\documentclass{jetpl}
\twocolumn \lat

\def\gmin{\mathcal{G}}

\def\vmaxi{V_{1}}
\def\smaxi{\mathcal{S}_{1}}

\title{The density of stationary points\\ in a high-dimensional
random energy landscape\\ and the onset of glassy behaviour}

\rtitle{The density of stationary points \ldots}

\author{Yan V Fyodorov$^{1,3}$, H-J Sommers$^{2}$, and Ian Williams$^{1}$}

\address{$^1$ School of Mathematical Sciences, University of
Nottingham, Nottingham NG72RD, England \\ $^2$ Fachbereich Physik,
Universit\"{a}t Duisburg-Essen, D-47048 Duisburg, Germany \\
$^3$ Petersburg Nuclear Physics Institute, Russian Academy of
Sciences, Gatchina, Leningrad region, 188300, Russia}

\abstract{We calculate the density of stationary points and minima
of a $N\gg 1$ dimensional Gaussian energy landscape. We use it to
show that the point of zero-temperature replica symmetry breaking in
the equilibrium statistical mechanics of a particle placed in such a
landscape in a spherical box of size $L=R\sqrt{N}$ corresponds to
the onset of exponential in $N$ growth of the cumulative number of
stationary points, but not necessarily the minima. For finite
temperatures we construct a simple variational upper bound on the
true free energy of the $R=\infty$ version of the problem and show
that this approximation is able to recover the position of the whole
de-Almeida-Thouless line.}

\PACS{05.40.-a, 75.10.Nr}

\begin{document}

\maketitle

 Statistical properties of stationary
points (extrema as well as saddles) of random high-dimensional
landscapes continue to attract considerable interest in the physical
and mathematical literature, both analytically and numerically.
Motivations come from fields of research as diverse as the study of
disordered and glassy systems and string theory - see
\cite{YFglass,glasses,string} and references therein. In particular,
considerable effort \cite{metast} has gone into trying to understand
what changes the intricate scenario of ergodicity breaking typical
for mean-field spin glass models implies in the statistics of the
associated free energy landscape (the so-called TAP\cite{TAP}
variational functional).

One of the simplest models known to display many non-trivial
hallmarks of glassy dynamics and thermodynamics is the
so-called toy model of a classical particle in a box of
characteristic size $L$ filled with a random potential. The
position of the particle is characterized by the coordinate vector
${\bf x}=(x_1,...,x_N),\,\, |{\bf x}|\le L$, and the standard
choice of the energy function is
\begin{equation}\label{fundef}
{\cal H}({\bf x})=\frac{\mu}{2}\sum_{k=1}^N x_k^2+V(x_1,...,x_N)
\end{equation}
with $\mu\ge 0$. The random Gaussian-distributed potential $V({\bf
x})$ is assumed to have zero mean, and its covariance is chosen to have a
form ensuring stationarity, isotropy, and a well-defined large
$N-$limit:
\begin{equation}\label{potential}
\left\langle V\left({\bf x}_1\right) \, V\left({\bf
x}_2\right)\right\rangle_V=N\,f\left(\frac{1}{2N}({\bf x}_1-{\bf
x}_2)^2\right)\,.
\end{equation}
Similarly, the natural scaling of the sample size is $L=R\sqrt{N}$
\cite{FS}. Here and henceforth the notation
$\left\langle\ldots\right\rangle_V$ stands for averaging over the
random potential $V({\bf x})$.

In the limit when both $N\to \infty$ and $R\to\infty$ the
thermodynamics of the model has enjoyed a long history of research
starting from early works\cite{MP},\cite{Engel}, and\cite{CKD}. A
detailed discussion of the limit $N\to \infty$ for the fixed box
size $R<\infty$ can be found in a recent paper\cite{FS}. The paper
\cite{CL} investigates the related problem for finite dimensions
$N<\infty$ in a broad physical context.

The calculation in \cite{YFglass} revealed that the replica
symmetry breaking (interpreted in the standard way as ergodicity
breaking) in the zero-temperature limit of the $R=\infty$ version
of the model is accompanied by the emergence of an exponentially
large total number of stationary points in the energy landscape.
As naively one may expect that the low-temperature thermodynamics
should be dominated by minima rather than by the totality of
stationary points, the issue deserves further investigation, in
particular in the general $R<\infty$ case.

The goal of the present paper is to compute the density of all
 stationary points as well as of only
minima at a given value of the potential, hence energy, in the
limit $N\gg 1$ (in the simplest case $N=1$ similar questions were
addressed in \cite{1D}). Our analysis reveals that generally the
domain of existence of the glassy phase with broken ergodicity (at
zero temperature $T$) turns out to be associated with the
existence of exponentially many stationary points in the energy
landscape, but not necessarily exponentially many minima.  In an
attempt to extend our considerations to finite temperatures we
also construct a simple variational functional providing an upper
bound on the true free energy of the $R=\infty$ version of the
problem. Surprisingly, counting stationary points in this
simple-minded approximation is able to capture such a nontrivial
feature as the precise position of the de-Almeida-Thouless line in
the whole $(\mu,T)$ plane.

Our aim is to calculate the mean probability density
$\rho_s(V,{\bf x})$ of the value $V$ of the potential $V({\bf x})$
taken over all stationary points of the landscape ${\cal H}({\bf
x})$ around the position ${\bf x}$, with the stationary points of
any index counted with equal weight. We will also be able to
address its counterpart, $\rho_m(V,{\bf x})$, where only minima of
${\cal H}({\bf x})$ are taken into account. It is convenient to
define the joint probability density ${\cal F}(V,{\bf x},\hat{K})$
of the scalar argument $V$, the $N-$component vector argument
${\bf x}$ and the $N\times N$ real symmetric matrix argument
$\hat{K}$ as
\begin{multline}\label{1}
{\cal F}(V,{\bf x},\hat{K})=\left\langle\delta\left(V-V({\bf
x})\right)\right.\\ \times \left.\delta\left(\mu{\bf
x}+\frac{\partial}{\partial {\bf x}} V({\bf
x})\right)\delta\left(\hat{K}-\frac{\partial^2}{\partial {\bf
x}\partial{\bf x}}V({\bf x})\right)\right\rangle_V.
\end{multline}
in terms of which the density $\rho_s(V,{\bf x})$ can be expressed
as\cite{YFglass}
\begin{equation}\label{2}
\rho_s(V,{\bf x})=\int |\det{\left(\mu
\hat{I}_N+\hat{K}\right)}|\,{\cal F}(V,{\bf x},\hat{K})\,d\hat{K}
\end{equation}
where $\hat{I}_N$ stands for $N\times N$ identity matrix. The
density of minima $\rho_m(V,{\bf x})$ is given by a similar
integral, but with the integrand containing the (matrix)
step-function factor $\theta\left(\mu \hat{I}_N+\hat{K}\right)$
ensuring that all the eigenvalues of the matrix $\mu
\hat{I}_N+\hat{K}$ are positive.

Introducing the Fourier integral representation for each of the
delta-functional measures (of scalar, vector, or matrix argument)
in Eq.(\ref{1}) facilitates performing the ensemble average
explicitly in view of the Gaussian nature of the potential
Eq.(\ref{fundef}) and its stationarity, Eq.(\ref{potential}).
After some straightforward but long manipulations we find:
\begin{multline}\label{8a}
 {\cal F}(V,{\bf x},\hat{K})\propto \exp\left\{\frac{\mu^2{\bf
x}^2}{2f'(0)}-\frac{V^2}{2Nf(0)}\right\}\\
\times \int_{-\infty}^{\infty}
e^{-\frac{N}{2}t^2-\frac{N}{4\mu_{cr}^2}\mbox{Tr}\left[\hat{K}-\left(gt+\frac{f'(0)}{f(0)}\,\frac{V}{N}\right)\hat{I}_N\right]^2}
\,dt
\end{multline}
Here, and henceforth, we systematically disregard various
multiplicative constant factors for the sake of brevity and use
the following notations:
\begin{equation}\label{9}
\mu_{cr}=\sqrt{f''(0)}\,, \quad g^2=f''(0)-\frac{f'(0)^2}{f(0)}\ge
0\,.
\end{equation}
Substituting Eq.(\ref{8a}) back into Eq.(\ref{2}), and introducing
the notations
$$\hat{H}=\hat{K}-\left(gt+\frac{f'(0)}{f(0)}\,\frac{V}{N}\right)\hat{I}_N
,\quad \mu_{ef}(t,V)=\mu+gt+\frac{f'(0)}{f(0)}\,\frac{V}{N}$$ we
find that the density function factorizes as
\begin{equation}\label{2a}
\rho_s(V,{\bf x})={\cal N}_s\,\rho_0({\bf x})\rho_s(V)\, ,
\end{equation}
where
\begin{equation}\label{2b}
\rho_0({\bf x})=e^{\frac{\mu^2{\bf x}^2}{2f'(0)}},\quad \rho_s(V)=
\,e^{-\frac{V^2}{2Nf(0)}}\int_{-\infty}^{\infty}
e^{-\frac{N}{2}t^2}\,{\cal D}_s(t,V)dt
\end{equation}
with
\begin{equation}\label{10}
{\cal D}_s(t,V)=\int |\det{\left(\mu_{ef}(t,V)
\hat{I}_N+\hat{H}\right)}|e^{-\frac{N}{4\mu_{cr}^2}\mbox{Tr}\hat{H}^2}\,\,d\hat{H}\,,
\end{equation}
and the factor ${\cal N}_s$ accounts for the overall
normalisation. Analogously, the density of minima $\rho_m(V,{\bf
x})$ is given by a very similar expression, the only difference
being that Eq.(\ref{10}) is replaced by
\begin{multline}\label{10a}
 \quad \quad {\cal D}_m(t,V)= \int e^{-\frac{N}{4\mu_{cr}^2}\mbox{Tr}\hat{H}^2}
 \\ \times \det{\left(\mu_{ef}(t,V)
\hat{I}_N+\hat{H}\right)}\theta\left(\mu_{ef}(t,V)
\hat{I}_N+\hat{H}\right)\,\,d\hat{H}.
\end{multline}

 Our main interest is to extract the leading
asymptotics of these expressions in the limit $N\gg 1$. To this
end, for the analysis of $\rho_s(V)$ we use the results from
\cite{YFglass}, where it was demonstrated that essentially
$D_s(t,V)$ can be replaced in the large $N$ limit with
\begin{equation}\label{11}
{\cal D}_s(t,V)\propto \exp{N\Phi\left(s\right)}, \quad
s=\mu_{ef}(t,V)/\mu_{cr}
\end{equation}
where the function $\Phi(s)=\Phi(-s)$ is given explicitly by
\begin{equation}\label{12a}
\Phi(s\ge
0)=\frac{s^2}{4}-\theta(s-2)\left[\frac{s\,\sqrt{s^2-4}}{4}-\ln{\left(\frac{s+\sqrt{s^2-4}}{2}\right)}\right]
\end{equation}
Perhaps, the shortest way to understand the above asymptotics is
to notice that actually,
\begin{equation}\label{12b}
\Phi[s]=\,\int_{-2}^2\ln{|s+\lambda|}\,\nu_{sc}(\lambda)\,d\lambda
,\quad \nu_{sc}(\lambda)=\frac{1}{2\pi}\sqrt{4-\lambda^2}
\end{equation}
where the integral is understood in the sense of the principal
value. The above formula can be verified by applying ideas from
statistical mechanics to the evaluation of $D_s(t,V)$. A rigorous
exposition of the method and further references can be found in
\cite{BPS}, and a related informal treatment is given in a recent
insightful paper \cite{DeanMaj}.

To ensure a well-defined large $N$ limit we rescale $V\to NV$, and
evaluating the integral over $t$ with the Laplace method we find
that $\rho_s(V)\propto \exp{[-N{\cal S}(V)]}$, with
\begin{equation}\label{13}
{\cal
S}(V)=\frac{V^2}{2f(0)}+\mbox{min}_{t}\left[\frac{t^2}{2}-\Phi\left(\mu_{ef}(t,V)/\mu_{cr}\right)\right]
\end{equation}
Correspondingly, in the large $N$ limit the density function
$\rho_s(V)$ displays a pronounced maximum around the most probable
value $V=V_*=\mbox{argmin}\,{\cal S}(V)$, which can be found from
the system of equations:
\begin{equation}\label{14}
V_*=\frac{f'(0)}{\mu_{cr}}\frac{d\Phi(s)}{ds}|_{s=s_*},\quad
t_*=\frac{g}{\mu_{cr}}\frac{d\Phi(s)}{ds}|_{s=s_*},
\end{equation}
 where we have used the notation
\begin{equation}\label{15}
s_*=[\mu+gt_*+\frac{f'(0)}{f(0)}\,V_*]/\mu_{cr}\equiv
[\mu+\frac{\mu_{cr}^2}{f'(0)}\,V_*]/\mu_{cr}\,,
\end{equation}
the second equality following from the obvious relation
$t_*=gV_*/f'(0)$ implied by Eq.(\ref{14}), and the definition
Eq.(\ref{9}). Moreover, for $s_*\ge 0$ Eq.(\ref{12b}) implies that
\begin{equation}\label{16}
\frac{d\Phi(s)}{ds}|_{s=s_*}=\frac{1}{2}\left[s_*-\theta(s_*-2)\sqrt{s_*^2-4}\right]
\end{equation}
Assuming first $0\le s_*\le 2$, the first of the equations (\ref{14})
together with Eqs.(\ref{15},\ref{16}) immediately yields $V_*=\mu
f'(0)/\mu_{cr}^2$ and $s_*=2\mu/\mu_{cr}$, the solution being
compatible with our assumption for $0\le \mu\le\mu_{cr}$.
On the other hand, assuming $s_*>2$ the first of the equations
(\ref{14}) solved together with Eq.(\ref{16}) yields the relation
\begin{equation}
s_*=\frac{\mu_{cr}}{f'(0)}V_*+\left(\frac{\mu_{cr}}{f'(0)}V_*\right)^{-1} ,
\end{equation}
which is consistent with our assumption. Substituting for $s_{*}$ from
Eq.(\ref{15}) one immediately finds $V_*=f'(0)/\mu$. Summarizing,
we see that the most probable value $V_*$ of the potential $V({\bf
x}))$ at a stationary point of the energy surface ${\cal H}({\bf
x})$ is given by
\begin{equation}\label{17}
V_*= \frac{f'(0)}{\mu}\,\theta( \mu-\mu_{cr})+ \mu
\frac{f'(0)}{\mu_{cr}^2}\,\theta(\mu_{cr}-\mu)
\end{equation}
The corresponding values of ${\cal S}(V_*)$ can be found from
Eq.(\ref{13}):
\begin{equation}\label{18}
{\cal S}(V_*)=\left\{\begin{array}{l}
-\frac{1}{2}+\ln{\left(\frac{\mu_{cr}}{\mu}\right)},\quad \mu\ge \mu_{cr}\\
-\frac{\mu^2}{2\mu_{cr}^2}\quad, \,\quad\quad\quad \mu<
\mu_{cr}\end{array}\right.
\end{equation}

The mean total number $\langle\#_s\rangle_V$ of stationary points
is obviously obtained from the density  $\rho_s(V,{\bf x})$ in
Eq.(\ref{2a}) by integrating it across the range of coordinates
inside the sample $|{\bf x}|\le L=R\sqrt{N}$ and over all possible
values of the potential $V$. The integral over $V$ is dominated by
the most probable value $V_*$ found above, whereas the coordinate
integration is easily performed by the Laplace method yielding
\begin{multline}\label{19}
\int_{|{\bf x}|\le R\sqrt{N}}\rho_0({\bf x})\,d{\bf x}\propto\\
\propto \exp{\frac{N}{2}\left\{\begin{array}{l}
\ln{\left(2\pi R_{\mu}^2\right)},\quad\quad\quad R>R_{\mu}=\frac{\sqrt{|f'(0)|}}{\mu}\\
\ln{\left(2\pi R^2\right)}+1-\frac{R^2}{R^2_{\mu}}\quad, \,\quad
R<R_{\mu}\end{array}\right\}}
\end{multline}
Collecting all exponential factors from Eqs.(\ref{18},\ref{19})
and also restoring the overall normalisation from the requirement
$\lim_{\mu\to\infty}\langle\#_s\rangle_V=1$, one finds that the
associated cumulative complexity of stationary points
$\Sigma_s(\mu,R)=\lim_{N\to
\infty}\frac{1}{N}\ln{\langle\#_s\rangle_V}$ is given by
\begin{multline}
\label{20} \Sigma_s=
\frac{1}{2}\theta{\left(R_{\mu}-R\right)}\left[\ln{\left(\frac{R^2}{R^2_{\mu}}
\right)}+1-\frac{R^2}{R^2_{\mu}}\right]\le 0\,
\end{multline}
as long as $\mu\ge \mu_{cr}$, whereas for $\mu<\mu_{cr}$ the
complexity $\Sigma_s(\mu,R)$ is given by
\begin{multline}
\label{21}\begin{array}{l}
\Sigma_s=\frac{1}{2}\left[-1+\frac{\mu^2}{\mu^2_{cr}}-\ln{\left(\frac{\mu^2}{\mu^2_{cr}}\right)}\right]>
0\quad\quad\mbox{for}\quad\quad R\ge R_{\mu}\\
\Sigma_s=\frac{1}{2}\left[\ln{\left(\frac{R^2}{R^2_{cr}}
\right)}+\frac{\mu^2}{\mu^2_{cr}}\left(1-\frac{R^2}{R^2_{cr}}\right)\right]\,,
\quad R<R_{\mu}\end{array}
\end{multline}
where introduced an important lengthscale, cf.\cite{FS}
\begin{equation}
R_{cr}=\sqrt{|f'(0)|/f''(0)}\,.
\end{equation}

The calculation of the density of minima $\rho_{m}(V)$ proceeds in
a similar way. The required large $N$ asymptotics of
$\mathcal{D}_{m}(t,V)$ in Eq.(\ref{10a}) can be extracted from the
results of a recent paper \cite{DeanMaj}, and is given by
\begin{equation}
 \mathcal{D}_{m}(t,V) \propto \exp\left(-\frac{N^{2}}{2}\theta(2-s)\gmin(s) + N {\cal \phi}(s)\right),
\end{equation}
where $s$ has the same meaning as in Eq.(\ref{11}), ${\cal
\phi}(s)$ is given by the right-hand side of Eq.(\ref{12a}) for $s
\geq 2$, and
\begin{multline}
\gmin(s) =  \frac{1}{216}\left[72s^2 - s^4 -
(s^{3}+30s)\sqrt{12+s^{2}} \right]-\\
-\log\left(\frac{s+\sqrt{s^2+12}}{6}\right)\,.
\end{multline}
  When calculating $\rho_m(V)$ we again rescale $V \to NV$,
subdivide the integration range into two pieces $s\in(-\infty,2]$
and $s\in [2,\infty)$ and evaluate two resulting integrals by the
Laplace method in the large $N$ limit. The first integral
$I_{1}(V)$ is dominated by the point where $\gmin(s)$ is
minimized, i.e. by the vicinity of $s = 2$, the endpoint of the
integration range,  where actually $\gmin = 0$. Remembering that
${\cal \phi}(2) = 1$, we find after some manipulations $I_{1}(V)
\propto \exp-N \mathcal{S}_{1}(V)$, where
\begin{multline}
\smaxi(V) = \frac{1}{2g^{2}f(0)}\left[\mu_{cr} V - f'(0)\left(2 -
\frac{\mu}{\mu_{cr}}\right) \right]^{2} +\\ + \frac{1}{2}\left(2 -
\frac{\mu}{\mu_{cr}}\right)^2-1.
\end{multline}
Evidently, $I_{1}(V)$ is strongly peaked about the value
$V_{1}$, where
\begin{equation}\label{minasym}
 \vmaxi = \frac{f'(0)}{\mu_{cr}}\left(2 - \frac{\mu}{\mu_{cr}} \right),
\quad
 \smaxi(\vmaxi) = \frac{1}{2}\left(2 - \frac{\mu}{\mu_{cr}}\right)^2-1.
\end{equation}

In the second integral, $I_{2}(V)$, the arguments leading to
Eqs.(\ref{17},\ref{18}) show that for any $\mu$ it is dominated by
the same value $V=V_{*}=f'(0)/\mu$ as in the $\mu>\mu_{cr}$ case
in (\ref{17}), and is given by $I_{2}(V) \propto \exp-N
\mathcal{S}(V_*)$ with
$\mathcal{S}(V_{*})=-1/2+\ln{(\mu_{cr}/\mu)}$ (see (\ref{18})). It
is easy to verify that $\smaxi(\vmaxi) > \mathcal{S}(V_{*})$ for
$\mu>\mu_{cr}$, and for $\mu<\mu_{cr}$ the inequality is reversed,
giving the leading large $N$ asymptotics as
\begin{equation}\label{minasym1}
\rho_{m}(V)\propto \left\{\begin{array}{c} \exp-N
\mathcal{S}(V_*),\quad \mu>\mu_{cr}\\ \exp-N
\smaxi(\vmaxi)\,,\quad \mu<\mu_{cr}
\end{array}\right.
\end{equation}

We thus find that the associated complexity of the minima, given
by $\Sigma_{m}(\mu,R)=\lim_{N\to
\infty}\frac{1}{N}\ln{\langle\#_m\rangle_V} $, coincides with
$\Sigma_{s}(\mu,R)$ in the range $\mu \geq \mu_{cr}$. For
$\mu<\mu_{cr}$ the counterpart of (\ref{21}) is however given by a
different expression:
\begin{multline} \label{mincomplexity}  \Sigma_{m}=\frac{1}{2}\left[1-\left(2 - \frac{\mu}{\mu_{cr}}
\right)^2-\ln{\left(\frac{\mu^2}{\mu^2_{cr}}\right)}\right]> 0
\end{multline}
as long as $R\ge R_{\mu}$, whereas for $R<R_{\mu}$ it is given by
\begin{multline}\label{mincomplexity1} \Sigma_{m}=
\frac{1}{2}\left[2-\frac{R^2}{R^2_{cr}}\frac{\mu^2}{\mu^2_{cr}}+\ln{\left(\frac{R^2}{R^2_{cr}}\right)}-\left(2
- \frac{\mu}{\mu_{cr}} \right)^{2}\right]
\end{multline}

Let us first compare (\ref{21}) and
(\ref{mincomplexity},\ref{mincomplexity1}) in the infinite-sample
limit $R=\infty$ studied in \cite{YFglass}. In that case, we see
that {\it both} an exponentially large number of stationary
points, and exponentially many minima co-exist in precisely the
same interval $0<\mu\le\mu_{cr}$. Moreover, it is precisely the
same interval of the $\mu-$axis known to correspond at
zero-temperature to the glassy phase with broken
ergodicity\cite{MP,Engel,FS}.

Recent analysis of the general $R<\infty$ case of our model by the
replica trick \cite{FS} has revealed that at zero temperature the
phase with broken ergodicity covers the strip $R_{cr}<R <\infty
,\,0<\mu<\mu_{cr}$ in the $(R,\mu)$ plane. As is easy to infer
from the last line in expression Eq.(\ref{21}), the above strip is
exactly the domain supporting positive cumulative complexity
$\Sigma_{s}>0$. At the same time, analysis of
(\ref{mincomplexity}) shows that $\Sigma_{m}>0$ for any
$0<\mu<\mu_{cr}$ only when the sample radius satisfies
$R>R_m=e\,R_{cr}>R_{cr}$. On the other hand, if $R_{cr}<R<R_{m}$
then for the values of $\mu$ satisfying $0<\mu<\mu_{-}< \mu_{cr}$
with
\begin{equation}\label{last}
\mu_{-}=\frac{2\mu_{cr}}{1+R^2/R_{cr}^2}
\left[1-\sqrt{1-\frac{1+R^2/R_{cr}^2}{2}\ln{\frac{R_m}{R}}}\right]
\end{equation}
the complexity of minima $\Sigma_{m}$ is negative, although
 we have seen that $\Sigma_{s}>0$ in this range of parameters.
 We thus conclude that the replica symmetry - hence, ergodicity,-
 breaking implies that the landscape supports exponentially many
 stationary points, but not necessarily minima.

 It is natural to try to
understand whether similar relations exist at finite temperatures
for the {\it free} energy landscape. For the present model the
full TAP variational functional is as yet unknown, and
constructing it we consider as an interesting problem for a future
work. To this end, we propose below a simple variational
functional which provides only an upper bound for the true free
energy of our problem. Surprisingly enough, we find that such a
functional already contains enough information to reproduce the
position of the de-Almeida-Thouless line in the whole $(\mu,T)$
plane.

Concentrating, for simplicity, on the infinite sample $R=\infty$
case, we employ for our goal the
Gibbs-Bogoliubov-Feynman\cite{Ruelle} inequality:
\begin{equation}\label{22}
 F(H)\le F(H_a)+\left\langle H-H_a \right\rangle_{H_a}
\end{equation}
valid for any two Hamiltonians $H$ and $H_a$, with
$F(H)=-T\ln{\mbox{Tr}\, e^{-H/T}}$ and $F(H_a)=-T\ln{\mbox{Tr}\,
e^{-H_a/T}}$ standing for the corresponding free energies and
\begin{equation}\label{22a}
 \left\langle (\ldots) \right\rangle_{H_a}=\mbox{Tr}
\left[(\ldots) e^{-H_a/T}\right] /\mbox{Tr} e^{- H_a/T}\,.
\end{equation}
 The role of $H$ in our consideration will be played by
${\cal H}({\bf x})$ from Eq.(\ref{fundef}), and we choose $H_a$ in
the form $H_a({\bf x})=\frac{\mu}{2}({\bf m}-{\bf x})^2$ with the
$N-$component vector ${\bf m}$ playing the role of a variational
parameter.  For this choice, the inequality Eq.(\ref{22}) takes the
form
\begin{equation}\label{23}
F({\cal H})\le F_a({\bf m})=\frac{NT}{2}\ln{
\frac{\mu}{T}}+\frac{\mu}{2}{\bf m}^2+V_{a}({\bf m})
\end{equation}
where
\begin{equation}\label{23a}
\quad\quad\quad  V_{a}({\bf m})=\int  V\left({\bf m}+{\bf
y}\sqrt{\frac{T}{\mu}}\,\right)\,\exp\left\{-\frac{1}{2}{\bf
y}^2\right\}\, d{\bf y}\,,
\end{equation}
with $d{\bf y}=\prod_{i=1}^N \frac{d y_i}{\sqrt{2\pi}}$.
 The problem of finding the best possible
approximation to the true free energy in this class of trial
Hamiltonians obviously amounts to minimizing $F_a({\bf m})$ with
respect to the parameter ${\bf m}$. Note that $V_{a}({\bf m})$ is
obviously a random Gaussian function with zero mean. The
corresponding covariance function can be calculated using
Eq.(\ref{potential}) and Eq.(\ref{23a}) as
\begin{equation}\label{newpotential}
\left\langle V_a\left({\bf m}_1\right) \, V_a\left({\bf
m}_2\right)\right\rangle_V=N\,f_a\left(\frac{1}{2N}({\bf m}_1-{\bf
m}_2)^2\right),\quad
\end{equation}
where
\begin{equation}\label{24}
f_a\left({\bf b}^2\right)=\int  f\left[\left({\bf b}+{\bf
y}\sqrt{\frac{T}{\mu
N}}\,\right)^2\right]\,\exp\left\{-\frac{1}{2}{\bf
y}^2\right\}\,d{\bf y}\,,
\end{equation}
with the function $f$ taken from Eq.(\ref{potential}). The
invariance of the integrand in Eq.(\ref{24}) ensures that the last
expression indeed depends only on ${\bf b}^2$. Choosing ${\bf
b}=(|b|,0,\ldots,0)$, and rescaling ${\bf y}\to (y_1,\sqrt{N}{\bf
\tilde{y}})$, we can easily evaluate the integral over ${\bf
\tilde{y}}$ in the large $N$ limit by Laplace method. This
immediately leads to a very simple relation:
\begin{equation}\label{25}
\lim_{N\to \infty}f_a\left({\bf b}^2\right)=f\left({\bf
b}^2+\frac{T}{\mu}\right)
\end{equation}
Now it is evident that the free energy landscape represented by
the variational functional $F_a({\bf m})$ in Eq.({\ref{23}) is
very similar to that of the zero-temperature energy landscape
${\cal H}({\bf x})$ from Eq.(\ref{fundef}). In particular, the
functional $F_a({\bf m})$ should possess exponentially many
stationary points, and also minima, as long as
\begin{equation}\label{AT}
\mu^2<\mu^2_{cr}(T)=f''_{a}(0)\equiv
f''\left(\frac{T}{\mu}\right)\,.
\end{equation}
where we recall the definition Eq.(\ref{9}).
 The condition Eq.(\ref{AT}) coincides
precisely with one describing the position of the
de-Almeida-Thouless line in the $(\mu,T)$ plane for the present
model~\cite{MP,Engel,FS}.

When the major part of the calculations presented in this paper
were already completed, we became aware of a recent preprint by A
J Bray and D S Dean \cite{BrayDean} with a quite similar scope.
The authors restricted their presentation to the limiting case
$\mu=0$ of the same model, but were actually able to calculate
complexity of critical points with a given index. A characteristic
lengthscale $L_c$ appearing in their paper is simply related to
our $R_{m}=eR_{cr}$, see Eq.(\ref{last}), as $L_c=\sqrt{2\pi
e}\,\, R_{m}$, the extra factor accounting for the difference
between their cubic and our spherical shape of the sample in the
large $N$ limit.

{\bf Acknowledgements.} YVF acknowledges support of this project
by the Humboldt Foundation via the Bessel Award scheme. The
research in Nottingham was supported by EPSRC grant EP/C515056/1
"Random Matrices and Polynomials: a tool to understand
complexity", and in Duisburg by the SFB/TR 12 der DFG.


\begin{thebibliography}{99}
\bibitem{YFglass} Y V Fyodorov  {\it Phys. Rev. Lett.} {\bf 92}, Art. No.
240601 (2004); Erratum: {\it ibid.} {\bf 93}, Art. No. 149901
(2004) and  Acta Physica Polonica B, {\bf 36}, 2699 (2005)
\bibitem{glasses} K Broderix et al. {\it Phys. Rev. Lett.} {\bf 85} 5360 (2000); L
Angelani et al. {\it Phys. Rev. Lett.} {\bf 85} 5356 (2000); JPK
Doye, DJ Wales {\it J Chem Phys} {\bf 116} 3777 (2002);  TS
Grigera et al. {\it Phys. Rev. Lett.}  {\bf 88}, Art. No. 055502
(2002); F Sciortino {\it J Stat. Mech.: Th. $\&$ Exp.} Art. No.
P05015 (2005) ; TS Grigera {\it J. Chem. Phys. 124}, Art. No.
064502 (2006)
\bibitem{string}  M R Douglas, B Shiffman B, and S Zelditch {\it Comm. Math. Phys.} {\bf 252},
325 (2004) and {\it ibid} {\bf 265} 617 (2006)
\bibitem{metast} T. Aspelmeier, A. J.
Bray, and M. A. Moore {\it Phys. Rev. Lett.} {\bf 92}, 087203
(2004); A Crisanti et al.  {\it Phys. Rev. Lett.} {\bf 92}, 127203
(2004) ; G. Parisi and T Rizzo {\it J. Phys.A } {\bf 37}, 7979
(2004) ; A. Cavagna , I. Giardina and G. Parisi  {\it Phys. Rev.
Lett.} {\bf 92}, 120603 (2004);  T. Aspelmeier et al.  {\it Phys.
Rev. B}, {\bf 74}, 184411 (2006) ; M. M\"{u}ller, L. Leuzzi, and
A. Crisanti {\it Phys. Rev. B}, {\bf 74}, 134431 (2006)
\bibitem{TAP} D.J. Thouless, P.W. Anderson, and R.G. Palmer
{\it Philos. Mag.} {\bf 35}, 593 (1977)
\bibitem{FS} Y V Fyodorov and H-J Sommers {\it Nucl.Phys.B [FS]}
{\bf 764}, 128 (2007) [cond-mat/0610035]
\bibitem{MP} M.\, Mezard and G.\, Parisi {\it J.Phys.A:Math.Gen.} {\bf 23}, L1229 (1990);
 {\it J.Phys.I France} {\bf 1}, 809 (1991)
\bibitem{Engel} A.\, Engel {\it Nucl.Phys.B} {\bf 410}, 617 (1993) and {\it
J.Phys.Lett.} {\bf 46} L409 (1985)
\bibitem{CKD} L.F. \,Cugliandolo et al.
{\it Phys.Rev.Lett.} {\bf 76}, 2390 (1996) and {\it Phys.Rev. E}
{\bf 53}, 1525 (1996);
\bibitem{1D} A. \,Cavagna et al. {\it Phys.Rev.E} {\bf
59}, 2808 (1999)
\bibitem{CL} D.Carpentier, P. Le Doussal
{\it Phys.Rev.E} {\bf 63}, 026110 (2001)
\bibitem{BPS} A Boutet de Monvel, L Pastur and M Shcherbina {\it J. Stat. Phys.} {\bf 79} , 585 (1995)
\bibitem{DeanMaj} D S Dean , S N Majumdar
{\it Phys. Rev. Lett.} {\bf 97} Art. No. 160201 (2006)
\bibitem{Ruelle} D. Ruelle, {\it Statistical Mechanics: Rigorous
Results} (Benjamin, NY 1969)
\bibitem{BrayDean} A J Bray and D S Dean, e-preprint arXiv:cond-mat/0611023
\end{thebibliography}
\end{document}